\documentclass[10pt,conference]{IEEEtran}
\IEEEoverridecommandlockouts
\usepackage{cite}
\usepackage{amsmath,amssymb,amsfonts}
\usepackage{algorithmic}
\usepackage{graphicx}
\usepackage{textcomp}
\usepackage{xcolor}
\usepackage{tikz}
\usepackage{siunitx}
\usepackage[capitalise, noabbrev]{cleveref}

\definecolor{myorange}{HTML}{D6B36A}
\definecolor{mypeach}{HTML}{D49A84}
\definecolor{mypurple}{HTML}{8D7AB3}
\definecolor{myblue}{HTML}{4D8EAD}
\definecolor{mypink}{HTML}{D4877B}
\definecolor{mygreen}{HTML}{008f7a}
\definecolor{myteal}{HTML}{4e8397}

 \def\BibTeX{{\rm B\kern-.05em{\sc i\kern-.025em b}\kern-.08em
    T\kern-.1667em\lower.7ex\hbox{E}\kern-.125emX}}

\usepackage{acronym}
\acrodef{ADC}[ADC]{Analog to Digital Converter}
\acrodef{ADEXP}[AdExp-I\&F]{Adaptive-Exponential Integrate and Fire}
\acrodef{AdExp}[AdExp-IF]{Adaptive Exponential Integrate-and-Fire}
\acrodef{AE}[AE]{Address-Event}
\acrodef{AER}[AER]{Address-Event Representation}
\acrodef{AEX}[AEX]{AER EXtension board}
\acrodef{AFM}[AFM]{Atomic Force Microscope}
\acrodef{AGC}[AGC]{Automatic Gain Control}
\acrodef{AI}[AI]{Artificial Intelligence}
\acrodef{AMDA}[AMDA]{AER Motherboard with D/A converters}
\acrodef{ANN}[ANN]{Artificial Neural Network}
\acrodef{API}[API]{Application Programming Interface}
\acrodef{APMOM}[APMOM]{Alternate Polarity Metal On Metal}
\acrodef{ARM}[ARM]{Advanced RISC Machine}
\acrodef{ASIC}[ASIC]{Application Specific Integrated Circuit}
\acrodef{BCM}[BMC]{Bienenstock-Cooper-Munro}
\acrodef{BD}[BD]{Bundled Data}
\acrodef{BEOL}[BEOL]{Back-end of Line}
\acrodef{BG}[BG]{Bias Generator}
\acrodef{BL}[BL]{Bit Line}
\acrodef{BMI}[BMI]{Brain-Machince Interface}
\acrodef{BPTT}[BPTT]{Backpropagation Through Time}
\acrodef{BTB}[BTB]{band-to-band tunnelling}
\acrodef{CA}[CA]{Cortical Automaton}
\acrodef{CAD}[CAD]{Computer Aided Design}
\acrodef{CAM}[CAM]{Content Addressable Memory}
\acrodef{CAVIAR}[CAVIAR]{Convolution AER Vision Architecture for Real-Time}
\acrodef{CCN}[CCN]{Cooperative and Competitive Network}
\acrodef{CDR}[CDR]{Clock-Data Recovery}
\acrodef{CFC}[CFC]{Current to Frequency Converter}
\acrodef{CHP}[CHP]{Communicating Hardware Processes}
\acrodef{CIM}[CIM]{Compute In Memory}
\acrodef{CMIM}[CMIM]{Metal-insulator-metal Capacitor}
\acrodef{CML}[CML]{Current Mode Logic}
\acrodef{CMOL}[CMOL]{Hybrid CMOS nanoelectronic circuits}
\acrodef{CMOS}[CMOS]{Complementary Metal-Oxide-Semiconductor}
\acrodef{CNN}[CNN]{Convolutional Neural Network}
\acrodef{COTS}[COTS]{Commercial Off-The-Shelf}
\acrodef{CPG}[CPG]{Central Pattern Generator}
\acrodef{CPLD}[CPLD]{Complex Programmable Logic Device}
\acrodef{CPU}[CPU]{Central Processing Unit}
\acrodef{CSM}[CSM]{Cortical State Machine}
\acrodef{CSP}[CSP]{Constraint Satisfaction Problem}
\acrodef{CTXCTL}[CTXCTL]{CortexControl}
\acrodef{CV}[CV]{Coefficient of Variation}
\acrodef{DAC}[DAC]{Digital to Analog Converter}
\acrodef{DAS}[DAS]{Dynamic Auditory Sensor}
\acrodef{DAVIS}[DAVIS]{Dynamic and Active Pixel Vision Sensor}
\acrodef{DBN}[DBN]{Deep Belief Network}
\acrodef{DFA}[DFA]{Deterministic Finite Automaton}
\acrodef{DI}[DI]{delay insensitive}
\acrodef{DIBL}[DIBL]{drain-induced-barrier-lowering}
\acrodef{divmod3}[DIVMOD3]{divisibility of a number by three}
\acrodef{DMA}[DMA]{Direct Memory Access}
\acrodef{DNF}[DNF]{Dynamic Neural Field}
\acrodef{DNN}[DNN]{Deep Neural Network}
\acrodef{DOF}[DOF]{Degrees of Freedom}
\acrodef{DPE}[DPE]{Dynamic Parameter Estimation}
\acrodef{DPI}[DPI]{Differential Pair Integrator}
\acrodef{DR}[DR]{Dual Rail}
\acrodef{DRAM}[DRAM]{Dynamic Random Access Memory}
\acrodef{DRRZ}[DR-RZ]{Dual-Rail Return-to-Zero}
\acrodef{DSP}[DSP]{Digital Signal Processor}
\acrodef{DVS}[DVS]{Dynamic Vision Sensor}
\acrodef{DYNAP}[DYNAP]{Dynamic Neuromorphic Asynchronous Processor}
\acrodef{EBL}[EBL]{Electron Beam Lithography}
\acrodef{ECG}[ECG]{Electrocardiography}
\acrodef{EDVAC}[EDVAC]{Electronic Discrete Variable Automatic Computer}
\acrodef{EEG}[EEG]{Electroencephalography}
\acrodef{EIN}[EIN]{Excitatory-Inhibitory Network}
\acrodef{EM}[EM]{Expectation Maximization}
\acrodef{EMG}[EMG]{Electromyography}
\acrodef{eNVM}[eNMV]{Embedded Non-Volatile Memory}
\acrodef{EPSC}[EPSC]{Excitatory Post-Synaptic Current}
\acrodef{EPSP}[EPSP]{Excitatory Post-Synaptic Potential}
\acrodef{ES}[ES]{Evolutionary Strategies}
\acrodef{ESN}[ESN]{Echo state Network }
\acrodef{EZ}[EZ]{Epileptogenic Zone}
\acrodef{FDSOI}[FDSOI]{Fully-Depleted Silicon on Insulator}
\acrodef{FET}[FET]{Field-Effect Transistor}
\acrodef{FFT}[FFT]{Fast Fourier Transform}
\acrodef{FI}[F-I]{Frequency-Current}
\acrodef{fMRI}[fMRI]{functional Magnetic Resonance Imaging}
\acrodef{FPGA}[FPGA]{Field Programmable Gate Array}
\acrodef{FR}[FR]{Fast Ripple}
\acrodef{FSA}[FSA]{Finite State Automaton}
\acrodef{FSM}[FSM]{Finite State Machine}
\acrodef{GIDL}[GIDL]{gate-induced-drain-leakage}
\acrodef{GOPS}[GOPS]{Giga-Operations per Second}
\acrodef{GPU}[GPU]{Graphical Processing Unit}
\acrodef{GUI}[GUI]{Graphical User Interface}
\acrodef{HAL}[HAL]{Hardware Abstraction Layer}
\acrodef{HCS}[HCS]{High-Conductive State}
\acrodef{HFO}[HFO]{High Frequency Oscillation}
\acrodef{HH}[H\&H]{Hodgkin \& Huxley}
\acrodef{HMM}[HMM]{Hidden Markov Model}
\acrodef{HR}[HR]{Human Readable}
\acrodef{HRS}[HRS]{High-Resistive State}
\acrodef{HSE}[HSE]{Handshaking Expansion}
\acrodef{HW}[HW]{Hardware}
\acrodef{hWTA}[hWTA]{hard Winner-Take-All}
\acrodef{IC}[IC]{Integrated Circuit}
\acrodef{ICT}[ICT]{Information and Communication Technology}
\acrodef{IEEG}[iEEG]{intracranial electroencephalography}
\acrodef{IF}[I\&F]{Integrate-and-Fire}
\acrodef{IF2DWTA}[IF2DWTA]{Integrate \& Fire 2--Dimensional WTA}
\acrodef{IFSLWTA}[IFSLWTA]{Integrate \& Fire Stop Learning WTA}
\acrodef{IMC}[IMC]{In-Memory Computing}
\acrodef{IMU}[IMU]{Inertial Measurement Unit}
\acrodef{INCF}[INCF]{International Neuroinformatics Coordinating Facility}
\acrodef{INI}[INI]{Institute of Neuroinformatics}
\acrodef{IO}[I/O]{Input/Output}
\acrodef{IoT}[IoT]{Internet of Things}
\acrodef{IP}[IP]{Intellectual Property}
\acrodef{IPSC}[IPSC]{Inhibitory Post-Synaptic Current}
\acrodef{IPSP}[IPSP]{Inhibitory Post-Synaptic Potential}
\acrodef{ISI}[ISI]{Inter-Spike Interval}
\acrodef{JFLAP}[JFLAP]{Java - Formal Languages and Automata Package}
\acrodef{LCS}[LCS]{Low-Conductive State}
\acrodef{LEDR}[LEDR]{Level-Encoded Dual-Rail}
\acrodef{LFP}[LFP]{Local Field Potential}
\acrodef{LIF}[LIF]{Leaky Integrate and Fire}
\acrodef{LLC}[LLC]{Low Leakage Cell}
\acrodef{LLM}[LLM]{Large Language Model}
\acrodef{LNA}[LNA]{Low-Noise Amplifier}
\acrodef{LPF}[LPF]{Low Pass Filter}
\acrodef{LRS}[LRS]{Low-Resistive State}
\acrodef{LSM}[LSM]{Liquid State Machine}
\acrodef{LTD}[LTD]{Long Term Depression}
\acrodef{LTI}[LTI]{Linear Time-Invariant}
\acrodef{LTP}[LTP]{Long Term Potentiation}
\acrodef{LTU}[LTU]{Linear Threshold Unit}
\acrodef{LUT}[LUT]{Look-Up Table}
\acrodef{LVDS}[LVDS]{Low Voltage Differential Signaling}
\acrodef{ML}[ML]{Machine Learning}
\acrodef{MCMC}[MCMC]{Markov-Chain Monte Carlo}
\acrodef{MEMS}[MEMS]{Micro Electro Mechanical System}
\acrodef{MFR}[MFR]{Mean Firing Rate}
\acrodef{MIM}[MIM]{Metal Insulator Metal}
\acrodef{MLP}[MLP]{Multilayer Perceptron}
\acrodef{MOS}[MOS]{Metal Oxide Semiconductor}
\acrodef{MOSCAP}[MOSCAP]{Metal Oxide Semiconductor Capacitor}
\acrodef{MOSFET}[MOSFET]{Metal Oxide Semiconductor Field-Effect Transistor}
\acrodef{MRI}[MRI]{Magnetic Resonance Imaging}
\acrodef{ND}[ND]{Noise-Driven}
\acrodef{NDFSM}[NDFSM]{Non-deterministic Finite State Machine} 
\acrodef{NEF}[NEF]{Neural Engineering Framework}
\acrodef{NHML}[NHML]{Neuromorphic Hardware Mark-up Language}
\acrodef{NIL}[NIL]{Nano-Imprint Lithography}
\acrodef{NMDA}[NMDA]{N-Methyl-D-Aspartate}
\acrodef{NME}[NE]{Neuromorphic Engineering}
\acrodef{NN}[NN]{Neural Network}
\acrodef{NRZ}[NRZ]{Non-Return-to-Zero}
\acrodef{NSM}[NSM]{Neural State Machine}
\acrodef{OR}[OR]{Operating Room}
\acrodef{OTA}[OTA]{Operational Transconductance Amplifier}
\acrodef{PCB}[PCB]{Printed Circuit Board}
\acrodef{PCHB}[PCHB]{Pre-Charge Half-Buffer}
\acrodef{PCM}[PCM]{Phase Change Memory}
\acrodef{PE}[PE]{Processing Element}
\acrodef{PFA}[PFA]{Probabilistic Finite Automaton}
\acrodef{PFC}[PFC]{prefrontal cortex}
\acrodef{PFM}[PFM]{Pulse Frequency Modulation}
\acrodef{PR}[PR]{Production Rule}
\acrodef{PSC}[PSC]{Post-Synaptic Current}
\acrodef{PSP}[PSP]{Post-Synaptic Potential}
\acrodef{PSTH}[PSTH]{Peri-Stimulus Time Histogram}
\acrodef{PVT}[PVT]{Process, Voltage, Temprature}
\acrodef{QDI}[QDI]{Quasi Delay Insensitive}
\acrodef{R}[R]{Ripples}
\acrodef{RAM}[RAM]{Random Access Memory}
\acrodef{RDF}[RDF]{random dopant fluctuation}
\acrodef{RELU}[ReLu]{Rectified Linear Unit}
\acrodef{RL}[RL]{Reinforcement Learning}
\acrodef{RLS}[RLS]{Recursive Least-Squares}
\acrodef{RMS}[RMS]{Root Mean Squared}
\acrodef{RMSE}[RMSE]{Root Mean Squared-Error}
\acrodef{MSE}[MSE]{Mean Squared-Error}
\acrodef{KL}[KL]{Kullback-Leibler}
\acrodef{RNN}[RNN]{Recurrent Neural Networks}
\acrodef{ROLLS}[ROLLS]{Reconfigurable On-Line Learning Spiking}
\acrodef{RRAM}[RRAM]{Resistive Random Access Memory}
\acrodef{RSNN}[RSNN]{Recurrent Spiking Neural Network}
\acrodef{SAT}[SAT]{Boolean Satisfiability Problem}
\acrodef{SCX}[SCX]{Silicon CorteX}
\acrodef{SD}[SD]{Signal-Driven}
\acrodef{SEM}[SEM]{Spike-based Expectation Maximization}
\acrodef{SHD}[SHD]{Spiking Heidelberg Digit}
\acrodef{SL}[SL]{Source Line}
\acrodef{SLAM}[SLAM]{Simultaneous Localization and Mapping}
\acrodef{SNN}[SNN]{Spiking Neural Network}
\acrodef{SNR}[SNR]{Signal to Noise Ratio}
\acrodef{SOC}[SOC]{System-On-Chip}
\acrodef{SOI}[SOI]{Silicon on Insulator}
\acrodef{SP}[SP]{Separation Property}
\acrodef{SRAM}[SRAM]{Static Random Access Memory}
\acrodef{SRNN}[SRNN]{Spiking Recurrent Neural Network}
\acrodef{SRNN}[SRNN]{Spiking Recurrent Neural Networks}
\acrodef{SSC}[SSC]{Spiking Speech Command}
\acrodef{STD}[STD]{Short-Term Depression}
\acrodef{STDP}[STDP]{Spike-Timing Dependent Plasticity}
\acrodef{STP}[STP]{Short-Term Plasticity}
\acrodef{STT}[STT]{Spin-Transfer Torque}
\acrodef{STT-MRAM}[STT-MRAM]{Spin-Transfer Torque Magnetic Random Access Memory}
\acrodef{SW}[SW]{Software}
\acrodef{sWTA}[sWTA]{soft Winner-Take-All}
\acrodef{TCAM}[TCAM]{Ternary Content-Addressable Memory}
\acrodef{TFT}[TFT]{Thin Film Transistor}
\acrodef{TPU}[TPU]{Tensor Processing Unit}
\acrodef{USB}[USB]{Universal Serial Bus}
\acrodef{VHDL}[VHDL]{VHSIC Hardware Description Language}
\acrodef{VLSI}[VLSI]{Very Large Scale Integration}
\acrodef{VOR}[VOR]{Vestibulo-Ocular Reflex}
\acrodef{WCST}[WCST]{Wisconsin Card Sorting Test}
\acrodef{WL}[WL]{Word Line}
\acrodef{WTA}[WTA]{Winner-Take-All}
\acrodef{XML}[XML]{eXtensible Mark-up Language}
\acrodef{SAC}[SAC]{Soft Actor Critic}
\acrodef{PD}[PD]{Policy Distillation}
\acrodef{STE}[STE]{Straight-Through Estimator}
\acrodef{SOTA}[SOTA]{State of the Art}
\acrodef{SSM}[SSM]{State Space Model}
\acrodef{GRU}[GRU]{Gate Recurrent Unit}
\acrodef{LSTM}[LSTM]{Long Short Term Memory}
\acrodef{TCN}[TCN]{Temporal Convolutional Network}
\acrodef{FeFET}[FeFET]{Ferroelectric Field Effect Transistor}
\acrodef{JEPA}[JEPA]{Joint-Embedding Predictive Architectures}
\acrodef{RPL}[RPL]{Recurrent Predictive Learning}
\acrodef{CPC}[CPC]{Contrastive Predictive Coding}
\acrodef{RTRL}[RTRL]{Real-Time Recurrent Learning}
\acrodef{LRU}[LRU]{Linear Recurrent Unit}
\acrodef{NoC}[NoC]{Network-on-Chip}

\begin{document}

\title{Small-World Communication Fabrics for Neuromorphic Multicore-SoCs

\thanks{This work is supported by SNSF Starting Grant Project UNITE (TMSGI2-211461).}}

\author{\IEEEauthorblockN{
Sebastian Billaudelle\IEEEauthorrefmark{1},
Christian Metzner\IEEEauthorrefmark{1}, 
Jimmy Weber\IEEEauthorrefmark{1}, Zhe Su\IEEEauthorrefmark{1}, Chenxi Wen\IEEEauthorrefmark{1}, Siqi Liu\IEEEauthorrefmark{1}, \\
Laura Kriener\IEEEauthorrefmark{1}, Filippo Moro\IEEEauthorrefmark{1}, Giacomo Indiveri\IEEEauthorrefmark{1}, Melika Payvand\IEEEauthorrefmark{1}}

\IEEEauthorblockA{
\IEEEauthorrefmark{1}Institute of Neuroinformatics,
University of Zurich and ETH Zurich, Zurich, Switzerland  \\
Email: \{sebastian, melika\}@ini.uzh.ch }}

\maketitle

\thispagestyle{plain}
\pagestyle{plain}

\begin{abstract}
As neuromorphic systems scale beyond a single core, inter-core event communication can become a dominant contributor to memory footprint, latency, and energy consumption. Biological neural systems address a similar scaling challenge through small-world organization, combining dense local connectivity with sparse long-range projections. In this work, we compare two recent multicore neuromorphic systems implemented in the same 22-nm FDSOI technology and explicitly optimized for such connectivity. The first, NeoCorAl, uses an asynchronous packet-switched tree with hierarchical multicast, whereas the second, MOSAIC, employs an RRAM-based, circuit-switched two-dimensional mesh that performs routing in memory. We examine the resulting trade-offs in routing flexibility, hop count, memory requirements, multicast efficiency, and scalability. We further study how the relative efficiency of tree- and mesh-based routing depends on communication locality in spatially-embedded, random, and layered networks. Finally, we discuss routing-aware training as a means of jointly optimizing neural connectivity, task performance, and hardware mappability.
\end{abstract}


\section{Introduction}

\label{sec:intro}


The energy cost of computing increasingly lies in moving data rather than performing computation~\cite{horowitz_2014_vlsi}. 
Neuromorphic processors reduce this cost by distributing memory and computation across neural cores and communicating sparse events instead of dense numerical activations. However, the capacity of an individual core is limited by the size of its neuron and synapse arrays. Large neural networks must therefore be partitioned across multiple cores, and the cost of transmitting events between them can become a substantial fraction of the system area, energy, and latency~\cite{young2019review}.

Inter-core communication is often implemented using \ac{AER} in which the identity of a firing neuron is encoded into a digital event and routed toward its destination synapses~\cite{boahen_1998_aer}. Supporting arbitrary neuron-to-neuron connectivity is highly flexible, but requires routing and interconnect resources. These costs grow with network size and fan-out, motivating architectures that restrict supported connectivity or exploit structured communication. Typically, neuromorphic systems group neurons into cores with dense local connectivity and use a more constrained fabric for the sparser communication between cores~\cite{young2019review}.

This organization is consistent with biological neural systems, where connectivity is predominantly local and decreases with physical distance, resulting in sparse projections across distant regions. The resulting small-world organization combines highly connected local neighborhoods with sparse global integration~\cite{bullmore2009complex}. 
For electronics, the same structure is attractive for physical reasons, where local connections require shorter wires and fewer routing resources, whereas global communication incurs repeated switching, buffering, and memory-access costs. Locality can therefore be treated as an architectural principle for scalable neuromorphic systems~\cite{kudithipudi_etal2025_nmcscale}.

Existing neuromorphic processors realize this principle with different communication topologies. 
\textit{Tree-based and hierarchical fabrics} provide short paths across large systems and naturally support multicast by replicating events as they descend the hierarchy. 
Neurogrid, for example, connects chips through a binary multicast tree, while hierarchical \ac{AER} schemes extend local buses across steadily larger distances~\cite{neurogrid,merolla2014multicast,park_etal_2016_hiAER}. 
Such fabrics offers efficient global reach, but shared links toward upper levels can limit bandwidth and lead to congestion.

\textit{Mesh-based fabrics} forward events hop-by-hop through neighboring
routers in a regular physical topology. TrueNorth~\cite{truenorth} uses an
asynchronous two-dimensional (2D) mesh with destination-based dimension-order
routing, SpiNNaker~\cite{furber_etal2014_spinnaker1} uses a 2D
triangular torus with routing-table-based multicast, and
Loihi~\cite{davies_etal2018_loihi} uses an asynchronous 2D
mesh with dimension-order unicast routing and source-side replication for
multicast.
BrainScaleS-1 follows a more statically configured approach, establishing event paths through wafer-scale buses~\cite{schemmel_etal2010_bss1,schmidt2023cleanroom}. Meshes provide a regular physical implementation, distributed bandwidth, and efficient communication between nearby cores, but the number of traversed links increases with source--destination distance.

Many systems combine these approaches. DYNAP-SE and DYNAP-SE2 both use hierarchical routing within a chip and a two-dimensional inter-chip mesh, together with local tag-based broadcast~\cite{dynap_se1, richter2024dynap}. BrainScaleS-1 takes the opposing approach of combining the wafer-wide mesh with tree-based inter-wafer routing. Earlier \ac{AER} work also explored flat buses, broadcast grids, pre-structured interconnects, hierarchical networks, and router meshes, each offering a different trade-off between flexibility, bandwidth, latency, and routing-memory cost~\cite{zamarreno_etal_2012_aer}.

Despite their differences, these architectures share a motif: local communication within neural clusters is made inexpensive, while expensive communication over larger distances is progressively more constrained. The central design question is therefore how flexible the global connectivity should be and how communication cost should scale with distance.

In this work, we study this question through two recent multi-core neuromorphic systems designed explicitly for locally dense and globally sparse networks. Both are implemented in \SI{22}{\nano\meter} \ac{FDSOI} technology, allowing their architectural choices to be examined without conflating them with differences in the underlying technology. Nevertheless, they adopt fundamentally different communication fabrics:
\emph{NeoCorAL} uses an asynchronous packet-switched architecture with hardware support for tree-based multicast. 
Routing information is processed dynamically as events traverse the network, providing flexible one-to-many communication and short paths to distributed cores. \emph{MOSAIC} arranges neural and routing tiles in a two-dimensional mesh and uses \ac{RRAM} crossbars to statically configure event paths. It extends in-memory computing to \emph{in-memory routing}, replacing conventional routing \acp{LUT} with non-volatile circuit-switched connectivity~\cite{dalgaty_etal_2024_mosaic}.
We compare these approaches in terms of routing flexibility, hop count, memory footprint, and scalability. We then evaluate how their relative advantages depend on communication locality.
Finally, we discuss how routing constraints can be incorporated during network optimization, rather than treated only as post-training mapping limitations.



\section{Asynchronous packet-switched tree routing}

\label{sec:neocoral}

NeoCorAL is a mixed-signal multi-core neuromorphic processor whose inter-core spikes travel on an asynchronous, packet-switched tree \ac{NoC} realizing the \ac{AER} protocol.
Spikes are arbitrated, encoded, routed through asynchronous switches, and decoded at the destination to drive target synapses. 
The fabric consists of four levels: a cortex-inspired \emph{hierarchical routing model} for small-world connectivity, the \emph{core interface} (output arbitration/encoding and input routing memory), the \emph{data switch} (transport), and the \emph{addressing scheme} (routing).

\subsection{Cortex-inspired hierarchical routing}
\label{sec:smw}
The NeoCorAL \ac{NoC} follows cortical small-world connectivity, where long-range links become less frequent with distance~\cite{neocoral_swn}. Restricting the supported \acp{SNN} to this class shrinks the routing memory. 
NeoCorAL maps physical distance onto a router hierarchy: R0 broadcasts within a densely connected core (no intra-core specificity), R1 links $n$ local cores, and R2 links $n$ R1 routers. 
Higher routing levels expose progressively fewer connections, reducing routing-memory requirements.
Each router computes distance from local logic and memory and updates it in the packet as it ascends the tree, bounding the per-hop address space. A hardware-aware placement algorithm maps \acp{SNN} by extracting densely connected cliques into cores and allocating the numerous short-range connections before the sparse long-range ones. Since all cores at a level share a fixed address width, memory grows slowly with fan-in: a one-million-neuron network needs $\sim$67~Mbit, about $98\times$ less than DYNAP-SE~\cite{dynap_se1} and $307\times$ less than TrueNorth~\cite{akopyan2015truenorth}.

\subsection{Core interface: arbitration and CAM routing memory}
\label{sec:interface}
On the output side, a \emph{Hierarchical Arbiter Tree} (HAT) time-multiplexes parallel neuron outputs onto the \ac{NoC} ~\cite{neocoral_interface}. 
Four-input arbiters progressively merge spikes with low latency.
On the input path, an asynchronous \ac{CAM} routing \ac{LUT} filters events by single-cycle parallel search which is built on a NOR-type cell with current-race match-line sense amplifier.

\subsection{Asynchronous data-switch micro-architecture}
\label{sec:switch}
The transport layer uses an ultra-low-cost asynchronous switch designed for high-fan-out spiking traffic~\cite{neocoral_switch}. 
It uses a two-phase bundled-data protocol requiring one control transition per transfer, enabling high throughput without a clock and with minimal idle activity.
Its timing constraints are verified automatically using a commercial CAD flow.

Spikes are transmitted as compact \emph{single-flit packets}, with the header and payload combined into a single word. 
In each $N\times N$ switch, an Input Port Module (IPM) buffers the packet, reads its header, and sends a request to each selected output. 
Each Output Port Module (OPM) arbitrates competing requests, forwards the selected packet through the crossbar, and stores it in an output Mousetrap buffer.
For multicast, the crossbar replicates the packet to all selected OPMs, which can receive and forward it independently.


\subsection{Multicast addressing encoding scheme}
\label{sec:multicast}
The routing layer encodes target cores within the single-flit header while reusing the switch microarchitecture \cite{neocoral_multicast}.
NeoCorAL's \emph{Hierarchical Bit String} (HBS) scheme instead uses a $k$-bit mask per level to select the sub-regions that compose each multicast tree, so routing bits scale as $k /\log_2 k$ and addressing capability as $(2^k-1)^{1/\log_2 k}$. HBS maps directly onto the switch tree through a \emph{relative address with bit rotation} (MSB $=$ local branch), so every switch at a level shares identical, purely combinational routing logic. 


\section{RRAM-based circuit-switched mesh routing}

\label{sec:mosaic}

\begin{figure}[t]
    \centering

    \begin{tikzpicture}[remember picture]
        \clip (3.3, 3.7) rectangle ++(8.4, 4.4);

        \usetikzlibrary{arrows.meta}
        \usetikzlibrary{calc}
        
        \tikzset{
            >={Stealth[length=4pt, width=3pt]}
        }
        
        \foreach \i in {1,...,4} {
            \foreach \j in {1,...,3} {
                \node[draw,fill=lightgray!10!white,minimum width=2.4cm, minimum height=2.4cm] (tile-\i-\j) at (3*\i, 3*\j) {};
                \node[anchor=south east,inner sep=2pt] at (tile-\i-\j.south east) {\tiny core/tile};
                \node[draw=myblue,fill=myblue!30!white,minimum width=1.8cm, minimum height=0.7cm,yshift=0.55cm] (router-\i-\j) at (tile-\i-\j) {\tiny\color{myblue}{router}};
            }
        }

        \foreach \i in {1,...,3} {
            \pgfmathtruncatemacro{\next}{\i+1}
            \draw[<->,shorten <=0.5pt,shorten >=0.5pt,line width=1.5pt,myblue] (router-\i-2.east) -- (router-\next-2.west);
        }

        \foreach \i in {2,...,3} {
            \foreach \j in {1,...,2} {
                \pgfmathtruncatemacro{\next}{\j+1}
                \draw[<->,shorten <=0.5pt,shorten >=0.5pt,line width=1.5pt,myblue] (router-\i-\j.north) -- (router-\i-\next.south);
            }
        }

        \foreach \i in {1,...,4} {
            \foreach \j in {1,...,3} {
                \node[draw=myorange,fill=myorange!30!white,minimum width=1.0cm, minimum height=0.7cm,xshift=-0.4cm,yshift=-0.55cm] (weights-\i-\j) at (tile-\i-\j) {};
                \node[draw=myorange,fill=myorange!30!white,minimum width=0.5cm, minimum height=0.7cm,xshift= 0.3cm,yshift=-0.55cm] (weights-\i-\j-rec) at (tile-\i-\j) {};
                \node[draw=mypeach,fill=mypeach!30!white,minimum width=0.2cm, minimum height=0.7cm,xshift= 0.8cm,yshift=-0.55cm] (neurons-\i-\j) at (tile-\i-\j) {};

                \node[myorange,fill=myorange!30!white,xshift=0.25cm,inner sep=2pt] at (weights-\i-\j) {\tiny weights};
                \node[mypeach,rotate=90] at (neurons-\i-\j) {\tiny neurons};

                \draw[<-,shorten <=0.5pt,shorten >=0.5pt,line width=1.5pt,myblue] (weights-\i-\j.north) coordinate (tmp) -- (router-\i-\j.south -| tmp);
                \draw[<->,shorten <=0.5pt,shorten >=0.5pt,line width=1.5pt,mypink] (weights-\i-\j-rec.north) coordinate (tmp) -- (router-\i-\j.south -| tmp);
                \draw[shorten <=0.5pt,shorten >=0.5pt,line width=1.5pt,mypink] (neurons-\i-\j.north) coordinate (tmp) -- ($(tmp)!0.5!(tmp |- router-\i-\j.south)$) coordinate (tmp) -- (tmp -| weights-\i-\j-rec.north);

                \foreach \a in {-2,...,2} {
                    \draw[-{Stealth[length=3pt, width=2pt]},myorange,line width=1pt] ([yshift={\a*4pt}]weights-\i-\j-rec.east) coordinate (tmp) -- (tmp -| neurons-\i-\j.west);
                }
            }
        }
    \end{tikzpicture}

    \vspace{0.2cm}
    
    \begin{tikzpicture}[remember picture]
    \usetikzlibrary{decorations.pathreplacing}

    \tikzset{
        lrs-1-0/.style={mygreen},
        lrs-1-1/.style={mygreen},
        lrs-2-2/.style={mygreen}
    }
        
    \foreach \i in {0,...,3} {
    
        \foreach \j in {0,...,2} {
            \node[draw=mypeach,lrs-\i-\j/.try,minimum width=0.16cm,minimum height=0.35cm,inner sep=0pt] (r-\i-\j) at (\i,\j) {};

            \node[draw=mypeach,fill=mypeach,lrs-\i-\j/.try,minimum width=0.16cm,minimum height=0.1cm,anchor=south,inner sep=0pt] (r-fill-\i-\j) at (r-\i-\j.south) {};

            \draw[mypeach,lrs-\i-\j/.try] (r-fill-\i-\j.north) -- ++(0,0.05) -- ++(-0.05,0.0) -- ++(0.0,0.05) -- ++(0.1,0.0) -- ++(0.0,0.05) -- ++(-0.1,0.0) -- ++(0.0,0.05) -- ++(0.05,0.0) -- ++(0.0,0.05);

            \draw (r-\i-\j.south) -- ++(0,-0.1) -- ++(-0.1,0.0) -- ++(0.0,-0.2) -- ++(0.1,0.0) -- ++(0.0,-0.1) coordinate (tmp);
            \draw (tmp) ++(-0.05,0.0) -- ++(0.1,0.0);

            \draw (r-\i-\j.south) ++(0,-0.1) ++(-0.15,0.0) -- ++(0.0,-0.2) coordinate (tmp);
            \draw (tmp) ++(0,0.1) -- ++(-0.2,0.0) coordinate (wl-\i-\j);

            \draw (r-\i-\j.north) -- ++(0.0,0.1) coordinate (bl-\i-\j);
        }

        \draw (wl-\i-0) -- ++(0,2.9) coordinate (wl-\i);
        \foreach \j in {1,...,2} {
            \node[fill,circle,minimum width=1.5pt,inner sep=0pt] at (wl-\i-\j) {};
        }
    }

    \foreach \j in {0,...,2} {
        \draw (bl-0-\j) -- ++(4.0,0) -- ++(0.0,-0.2) coordinate (bl-\j);

        \foreach \i in {1,...,3} {
            \node[fill,circle,minimum width=1.5pt,inner sep=0pt] at (bl-\i-\j) {};
        }

        \node[anchor=south east, inner sep=0pt, yshift=0.24cm] at (bl-\j) {\tiny BL \j};

        \node[draw,anchor=north,minimum width=0.9cm,minimum height=0.5cm] (readout-\j) at (bl-\j) {\tiny readout};

        \draw (readout-\j.east) -- ++(0.5,0.0) coordinate (tmp);

        \draw (tmp) -- ++(0.0,0.25) -- ++(0.5,-0.25) -- ++(-0.5,-0.25) -- (tmp);

        \draw (tmp) ++(0.05,-0.10) -- ++(0.05,0.0) -- ++(0.08,0.2) -- ++(0.05,0.0);
        \draw (tmp) ++(0.05,0.0) -- ++(0.05,0.0) ++(0.08,0.0) -- ++(0.05,0.0);

        \draw[-stealth] (tmp) ++(0.5,0.0) -- ++(0.5,0.0) coordinate (out-\j);
    }

    \foreach \i in {0,...,3} {
        \node[anchor=west, inner sep=0pt, xshift=0.04cm, yshift=-0.03cm] at (wl-\i) {\tiny WL \i};
        
        \draw (wl-\i) -- ++(0,0.1) -- ++(0.15,0.3) -- ++(-0.3,0.0) -- ++(0.15, -0.3);

        \draw (wl-\i) ++(0,0.4) -- ++(0.0,0.1) coordinate (in-\i);
    }

    \draw [decorate, decoration={brace}] ([xshift=0.7cm,yshift=0.2cm]out-2) -- ([xshift=0.7cm,yshift=-0.2cm]out-0) node[pos=0.5,align=center,anchor=north,rotate=90,inner sep=1em] {\footnotesize to neighbor cores \& \\ \footnotesize weight crossbar};

    \draw [decorate, decoration={brace}] ([xshift=-0.2cm,yshift=0.5cm]in-0) -- ([xshift=0.2cm,yshift=0.5cm]in-3) node[pos=0.5,align=center,anchor=south,inner sep=0.5em] {\footnotesize from neurons \& \\ \footnotesize neighbor cores};

    \draw[line width=1pt,mypurple] (in-1) ++(-0.15,0.15) -- ++(0.1,0.0) -- ++(0.0,0.2) -- ++(0.1,0.0) -- ++(0.0,-0.2) -- ++(0.1,0.0);
    \draw[line width=1pt,myteal] (in-2) ++(-0.15,0.15) -- ++(0.1,0.0) -- ++(0.0,0.2) -- ++(0.1,0.0) -- ++(0.0,-0.2) -- ++(0.1,0.0);

    \draw[line width=1pt,mypurple] (out-0) ++(0.15,-0.1) -- ++(0.1,0.0) -- ++(0.0,0.2) -- ++(0.1,0.0) -- ++(0.0,-0.2) -- ++(0.1,0.0);
    \draw[line width=1pt,mypurple] (out-1) ++(0.15,-0.1) -- ++(0.1,0.0) -- ++(0.0,0.2) -- ++(0.1,0.0) -- ++(0.0,-0.2) -- ++(0.1,0.0);
    \draw[line width=1pt,myteal] (out-2) ++(0.15,-0.1) -- ++(0.1,0.0) -- ++(0.0,0.2) -- ++(0.1,0.0) -- ++(0.0,-0.2) -- ++(0.1,0.0);
\end{tikzpicture}

    \begin{tikzpicture}[remember picture,overlay]
        \draw[shorten <=0.1cm,shorten >=0.1cm,white,line width=1.5mm,line cap=round] (router-2-2.south west) -- ++(-1.8,-3.5);
        \draw[shorten <=0.1cm,shorten >=0.1cm,white,line width=1.5mm,line cap=round] (router-2-2.south east) -- ++( 2.5,-3.5);
        \draw[shorten <=0.1cm,shorten >=0.1cm,myblue,line width=0.5mm,line cap=round] (router-2-2.south west) -- ++(-1.8,-3.5);
        \draw[shorten <=0.1cm,shorten >=0.1cm,myblue,line width=0.5mm,line cap=round] (router-2-2.south east) -- ++( 2.5,-3.5);
    \end{tikzpicture}
    \vspace{-0.5cm}
    
    \caption{
        The MOSAIC multi-core architecture.
        Dense local connectivity is augmented by a more constraint mesh-based routing for long-range connections.
        Both weight and routing crossbars are based on non-volatile memory devices, implementing in-memory computing and routing paradigms.
    }
    \label{fig:mosaic}
\end{figure}
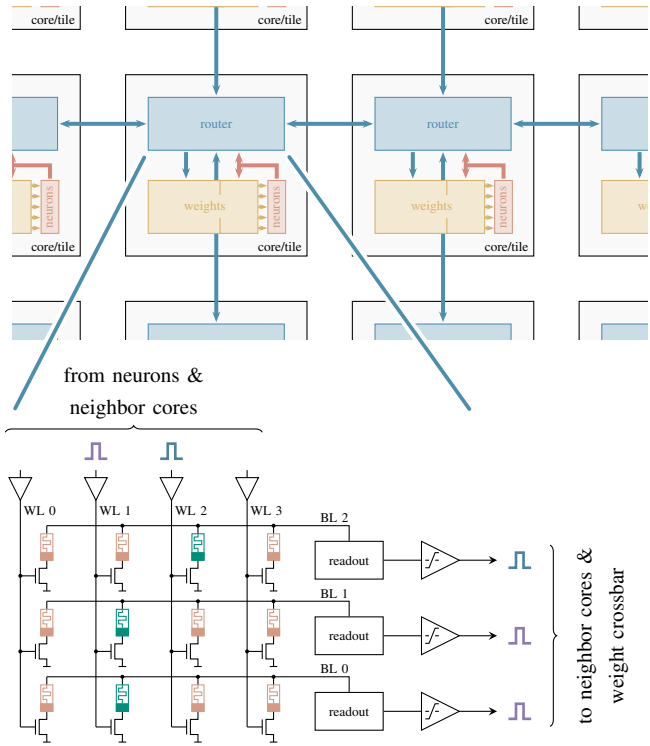

The MOSAIC architecture is likewise optimized for small-world network topologies, but -- despite all commonalities -- differs fundamentally from the system described in \cref{sec:neocoral}.
MOSAIC relies heavily on embedded non-volatile memory (eNVM) elements for both neural computation and on-chip event routing.
Going beyond traditional neuromorphic in-memory computing approaches, it extends the concept of in-memory computing to event routing.
This choice of technology has strongly influenced the overall design, and this section discusses the resulting design decisions and trade-offs.

\subsection{The MOSAIC architecture}

MOSAIC is designed as a highly scalable architecture featuring a homogeneous mesh of neuromorphic cores (\cref{fig:mosaic}).
Each core comprises \ac{LIF} neurons, which are implemented in pure \ac{CMOS} circuits, as well as a weight crossbar based on \ac{RRAM} devices for weight storage and the computation of input activations.
This crossbar is separated into two dedicated regions: a first block of synapses, each corresponding to two 1T1R branches to allow for positive as well as negative weights, realizes local all-to-all connectivity.
The second region receives inputs delivered by the core's router.
In addition to these dense core-internal recurrent connections, neurons can therefore receive input from additional sources, typically corresponding to longer-range inter-core projections.

\subsection{\ac{RRAM}-based in-memory routing}

Cores exchange events with their neighbors in all four cardinal directions, thereby forming a regular 2D mesh  (\cref{fig:mosaic}).
Communication routes are programmed statically and are inherently non-volatile because they are stored in \ac{RRAM} devices.
Consequently, the system can be suspended without losing its network configuration and requires neither weight nor route reloading after power-up.
The resulting communication scheme can therefore be characterized as non-volatile, statically configured circuit-switched routing.
Routes may connect neurons in directly adjacent cores or bypass intermediate cores to realize longer-range projections.
Moreover, routes can branch, enabling individual events to be delivered to multiple destinations and thus supporting multicast communication.

The router itself is implemented as an \ac{RRAM}-based switch matrix.
The crossbar's wordlines serve as inputs, and incoming events activate the corresponding wordlines and selected \ac{RRAM} devices.
A device programmed to the low-resistance state represents an active route, whereas a device in the high-resistance state represents an inactive connection.
The crossbar's bitlines represent the outgoing routes, and an outgoing event is triggered when increased conductance is detected on a bitline.
This approach requires a careful balance among device characteristics, crossbar sizes, and event statistics to avoid spurious events and spike loss.
Merging route memory with the switch implementation itself eliminates memory transfers and thus enables energy-efficient in-memory routing.
Without additional flow control and tagging, however, the switch matrices do not support many-to-one projections, which could result in conflicts between coincident events.
While the resulting one-hot encoding inherently decreases memory utilization and causes routing resources to scale linearly with the number of potential routes, this disadvantage is partly offset by the high density of the \ac{RRAM}-based approach.
Further improving memory utilization, MOSAIC employs buses with a \SI{4}{bit} source-address representation, reducing the required crossbar size by a factor of 16.

\begin{figure*}[t!]
    \centering

    \includegraphics[width=0.9\textwidth]{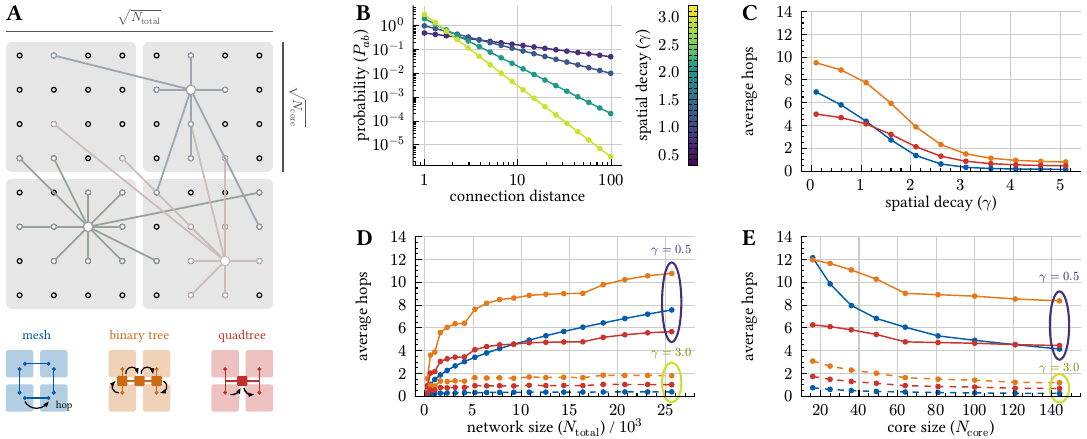}
    \vspace{-0.3cm}
    
    \caption{
        Comparison of inter-core routing architectures on spatially-embedded small-world network graphs.
        \textbf{A} Illustration of a spatially-embedded small-world network on a 2D grid of neurons.
        The connectivity of three units is exemplarily highlighted.
        This logical network can be mapped to a virtual multi-core system by partitioning the 2D neuronal grid into cores.
        These cores can be outfitted with varying routing schemes, e.g., a mesh or tree-based architectures.
        \textbf{B} A Kleinberg network's connectivity profile can be tuned via the spatial decay parameter $\gamma$.
        \textbf{C} A router's efficiency, here represented by the average number of hops per connection, depends especially on the density of long-range connectivity. Mesh routing excels at more local networks (high $\gamma$), while quadtree architectures are better suited for networks with a higher degree of global connectivity (low $\gamma$).
        \textbf{D} High-$\gamma$ networks scale well with growing size $N_\text{total}$ and allow mesh fabrics to exploit their efficient one-hop routing across neighboring cores.
        Quadtree routing becomes more efficient for Globally dense networks (low $\gamma$) with increasing connection distances.
        In that case, 
        \textbf{E} Mesh routing profits from larger cores, effectively reducing the distance (in hops), while a quadtree topology can cope with smaller cores more efficiently. Average fanout per neuron is fixed to $8$ across all sweeps.
        \vspace{-0.3cm}
    }
    \label{fig:routing_comparison}
\end{figure*}


The \ac{RRAM}-based in-memory routing concept is compatible with both asynchronous and synchronous communication paradigms.
The current implementation is fully synchronous, with each router hop introducing one clock cycle of latency.
An asynchronous routing scheme is equally feasible and was, in fact, adopted in the first MOSAIC prototype~\cite{dalgaty_etal_2024_mosaic}.
However, \ac{RRAM} crossbars and their readout circuitry are inherently analog and require well-defined pulses to ensure reliable read and programming operations.
Consequently, asynchronous communication would still require local pulse-generation circuitry at each routing stage.
The associated area and energy overhead would substantially reduce the latency and efficiency advantages typically attributed to asynchronous communication, making a synchronous implementation the more favorable trade-off in the present architecture.

%

\section{Efficiency of mesh- and tree-based routing}
\label{sec:comparison}

NeoCorAL and MOSAIC were both designed around a small-world layout.
Yet they adopt inherently different approaches for inter-core event routing, resulting in preferences for different network topologies.
In the following, we compare binary- and quadtree as well as mesh routing fabrics (\cref{fig:routing_comparison}A, bottom) in terms of routing cost across three network families: spatially embedded small-world networks, random networks, and feedforward layered networks.
We measure routing cost in hops, defined as the number of edges that need to be traversed across the routing fabric reaching from a source to a destination core, and normalize this figure by the total number of connections.
For each network and fabric, neurons are mapped onto cores to minimize routing cost~\cite{weber_etal_2023_gmap}.

\subsection{Spatially embedded small-world networks}

Spatially-embedded \textit{navigable} small-world connectivity is characterized by a connection probability that decays as a power law with the physical distance between nodes \cite{kleinberg_navigation_2000}.

We compare the average number of routing hops required to navigate such spatially-embedded small-world graphs mapped onto a virtual multi-core system (\cref{fig:routing_comparison}A). 
Following \cite{kleinberg_navigation_2000}, each neuron of the system is assigned a location on a 2D grid and connected to its nearest neighbors. 
A fixed number of long-range fan-out connections are generated per neuron with probability $P_{ab}\propto d_{ab}^{-\gamma}$
where $P_{ab}$ is the probability of connecting neurons $a$ and $b$ and $d_{ab}$ is the Euclidean distance between them on the 2D spatial embedding grid.
The clustering exponent $\gamma$ controls the trade-off between spatial locality and long-range communication across the small-world network, with higher $\gamma$ favoring more local connections while lower $\gamma$ leads to spatially distributed networks (\cref{fig:routing_comparison}B). 

\Cref{fig:routing_comparison}C indicates that the average number of routing hops required to navigate a mapped small-world network decreases monotonically with the clustering exponent $\gamma$ for all three routing architectures, provided a fixed number of total neurons ($N_\text{total}\approx 16,000$) and neurons per core ($N_\text{core}=64$), as well as a fixed average fanout of $8$.
In general, the binary tree requires more average routing hops than both mesh and quadtree.
At low $\gamma$, local connections dominate the network and the mesh benefits from the fact that neighboring cores are only one hop away while trees may need to navigate several levels of the hierarchy to connect neurons in cores that are spatially close but on different subtrees.
However, low $\gamma$ networks that arendistributed widely across the 2D grid favor the quadtree. 


In \cref{fig:routing_comparison}D, we observe that tree routing scales more favorably than a mesh when increasing the network size $N_\text{total}$ ($N_\text{core}$ fixed to $64$, average fanout of $8$) -- as long as $\gamma$ is sufficiently low.
However, increasing the number of neurons per core reduces the mesh routing cost even for a low $\gamma$ by decreasing the number of inter-core hops,  as shown in \cref{fig:routing_comparison}E where $N_\text{core}$ is swept while fixing $N_\text{total}\approx 16,000$.
This indicates that the range of spatially-embedded networks for which tree routing is advantageous decreases as more neurons are mapped to each core.

Overall, we find that mesh routing architectures shine on highly localized networks, while tree routing is best suited for larger networks that are more widely distributed across cores, especially in the case of relatively small cores.

\subsection{Random networks}

In contrast to small-world networks, fully random populations lack any spatial structure and their connectivity is often instead governed by a globally uniform probability density.
We thus consider them as a benchmark for how mesh and tree routing behave once spatial locality can no longer be exploited.

\begin{figure}[t]
\centering
\includegraphics[width=0.9\columnwidth]{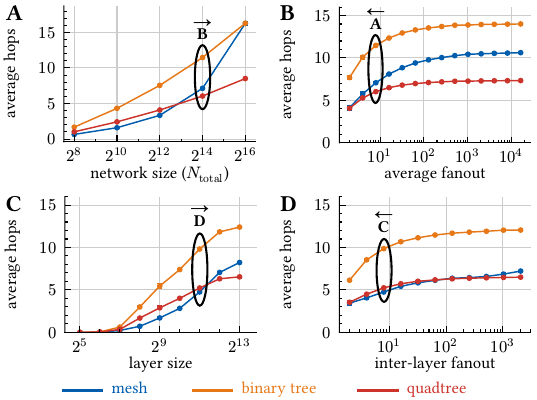}
\vspace{-0.35cm}
\caption{
    Comparison of inter-core routing cost for random (A, B) and layered feedforward networks (C,D), with $N_\text{core}=64$ and $N_\text{total}=16{,}384$ unless specified otherwise.
    Panels A and B as well as C and D represent two cuts through 2D parameter spaces; the fixed parameter is marked accordingly in the partner graph.
    \textbf{A} In random networks, the average number of hops per connection increases with growing network size $N_\text{total}$ (average fan-out fixed at $8$).
    Mesh routing becomes increasingly inefficient for larger networks.
    \textbf{B} Quadtree routing excels with an increasing fan-out and an increasing number of long-range connections.
    \textbf{C} Mesh routing maintains an advantage on feedforward networks with small- to medium-sized layers (inter-layer sparsity fixed at $2^{-8}$).
    \textbf{D} Mesh and quadtree routing fabrics exhibit a similar efficiency, independent of the inter-layer sparsity (8 layers of size $2048$).
}
\label{fig:comparison-random-layered}
\end{figure}

\Cref{fig:comparison-random-layered}A shows the average routing hops versus network size ($N_\text{total}$) at fixed average fan-out and $N_\text{core} = 64$.
For small networks, mesh routing benefits from the fact that most inter-core connections are between adjacent cores and outperforms both tree routers.
The quadtree becomes advantageous at larger network sizes, mirroring the low-$\gamma$ regime identified in \cref{fig:routing_comparison}D.
\Cref{fig:comparison-random-layered}B shows routing cost as a function of average fan-out, while keeping network size and thus the number of cores constant.
Here, quadtree scales most favorably.

Overall, random connectivity prevents mesh routing from profiting from strong locality, and does so more severely as the network grows: for small or sparse random networks all three fabrics are comparable, but for large or dense ones, the quadtree consistently offers lower routing costs.

\subsection{Feedforward layered networks}

Especially in machine learning applications, networks are often highly structured and directed.
Feedforward networks feature dense projections between consecutive layers but disregard layer-internal recurrence.
In contrast to small-world or fully random networks, placement algorithms can exploit the directionality for optimized routing.

In \cref{fig:comparison-random-layered}C, we consider a network of constant size and inter-layer sparsity.
For small layer size, connections remain mostly local and typically span only neighboring cores.
With growing layer sizes, layers begin to span multiple cores and the resulting placement can no longer fully exploit the directional data flow, thus incurring longer routes and a higher routing cost.
In this case, tree routers become more efficient, while mesh fabrics are optimal as long as communication remains mostly local.
In \cref{fig:comparison-random-layered}D, the number of layers is fixed at $8$, resulting in layers of size 2048, and the average inter-layer fan-out is swept.
Mesh and quadtree fabrics perform very similarly and remain more efficient than binary tree routing -- unlike in the random-network case (\cref{fig:comparison-random-layered}B) where the quadtree architecture is the best performing at high fan-out.

Preserving the structure and directionality of layered networks when mapping to a multi-core system is crucial to keep connectivity local.
Only then does mesh routing fully exploit the efficiency of communicating between adjacent cores.

\section{Routing-aware neural network training}
As shown in \cref{sec:comparison}, increasing spatial locality in small-world connected networks reduces the average number of routing hops required by the networks considered here. However, these gains are only realized if the neural network can be placed and routed efficiently on the hardware. The required resources depend not only on the number of connections, but also on where neurons are placed and how far their connections reach. 
A network may therefore satisfy the overall parameter budget yet use the routing resources inefficiently, or fail to map altogether because the connectivity supported by particular neuron or routing tiles is exceeded. Constraint-aware network optimization is thus central to exploiting the efficiency of a small-world communication fabric.
Common approaches to this problem face significant trade-offs: mapping only topologies that fit the hardware restrict adaptability, evolutionary architecture searches incur high computational costs, and gradient-based methods struggle with discrete, non-differentiable mapping and routing operations.


The routing-aware method in~\cite{weber2025hardware} adapts DeepR~\cite{bellec_etal_2017_deepR} to jointly optimize task performance and hardware mappability for MOSAIC. During training, an $\ell_1$ regularizer drives weak weights to zero, after which low-magnitude connections are pruned and reassigned while preserving a hardware-compatible distribution of connections across routing distances. This allows the network topology to adapt to the task without exceeding neuron- and routing-tile resources. Avoiding routing at every training step, the distance-dependent connectivity profile is used as an efficient proxy for mappability.

\begin{figure}[t]
\centering
\includegraphics[width=0.92\columnwidth]{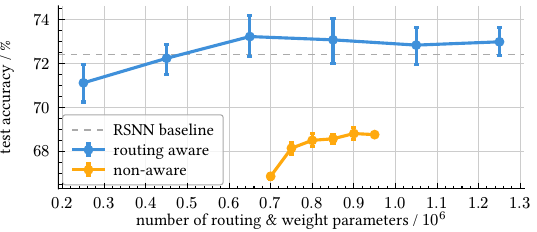}
\vspace{-0.35cm}
\caption{Test accuracy on the SHD data set as a function of the routing and weight memory count in the MOSAIC architecture \cite{dalgaty_etal_2024_mosaic}, trained with and without the routing-awareness, compared against the original vanilla Recurrent Spiking Neural Network (RSNN) results~\cite{cramer2020heidelberg}. Adapted from \cite{weber2025hardware}.
\vspace{-0cm}
}
\label{fig:deep_r_results}
\end{figure}

As shown in \cref{fig:deep_r_results}, applying this method to the Spiking Heidelberg Digits (SHD) task produces networks that are directly mappable onto MOSAIC. Compared with training a fixed hardware-compatible connectivity, routing-aware optimization achieves approximately five percentage points higher accuracy for the same memory budget and requires about one order of magnitude less memory at iso-accuracy~\cite{weber2025hardware}.

\section{Discussion}
Our comparison shows that small-world connectivity does not imply a single optimal routing topology. Mesh-based fabrics efficiently exploit strongly local traffic, whereas hierarchical trees provide shorter routes when communication is less local or spans larger systems. The appropriate fabric therefore depends on the spatial statistics of the target networks.

At the hardware level, the central challenge is to retain sufficient routing flexibility without recreating an expensive global communication fabric. Packet-switched trees provide flexible multicast and efficient global reach, but typically require routing tables and arbitration, and may experience congestion at higher levels of the hierarchy. In-memory circuit-switched meshes remove memory transfer and favor regular, local communication, but impose stronger constraints on route configuration and supported traffic. 

These constraints also create an algorithmic challenge. The benefits of a locality-optimized fabric are only realized when neural connectivity, placement, and routing are optimized together. As demonstrated by routing-aware training, two networks with the same number of parameters can require substantially different routing resources depending on how their connections are distributed in space. Compilation and learning should therefore be treated as part of the architectural design process. More broadly, locality can serve as an inductive bias toward modular computation, in which locally connected subnetworks specialize and communicate through a smaller number of structured long-range links.

Three-dimensional integration could further reshape these trade-offs. \ac{BEOL} memory integrated above logic can store synaptic and routing state close to computation, while sequential 3D integration can provide dense vertical connections between nearby neural and routing layers. At larger scales, hybrid bonding and through-silicon vias could connect clusters through a smaller number of long-range links~\cite{vianello_payvand_2024_3D}. This would naturally match small-world connectivity: dense, inexpensive communication within local three-dimensional neighborhoods and sparse communication between them. However, vertical interconnects, thermal budgets, and heterogeneous integration remain important constraints, making hardware-algorithm co-design essential.

In conclusion, communication efficiency in multi-core neuromorphic systems is not determined by routing topology alone. It emerges from the interaction between neural connectivity, physical communication fabric, placement and routing, and network optimization. Small-world organizations provides a common principle for this co-design, while mesh, tree, and future 3D fabrics offer complementary ways of realizing it.

\bibliography{bib/eisbib_hw,bib/eisbib_algo,bib/references}
\bibliographystyle{unsrt}

\end{document}